# Neutrino Long-Baseline Experiments and Nuclear Physics

Ulrich Mosel

Institut für Theoretische Physik

Universität Giessen

D-35392 Giessen, Germany

The Standard Model (SM) of particle physics has had its triumphant moment when in 2012 the Higgs particle was discovered at the LHC. Since then the research program at CERN has focused on finding new physics beyond the standard model (BSM). Since even the recent experimental runs did not yield any obvious new particles or phenomena the emphasis now is on precision physics. One prerequisite for these BSM precision studies is the remarkable beam precision and stability at LHC. The beam energy is known to better than 0.1% and its changes during a run are within about 0.02%. Enormous technical expertise, equipment and work went into reaching this accuracy. Up to today, however, no sign of BSM physics has been sighted.

It is, therefore, amusing to note that so far, the only indication of physics beyond the Standard Model has come from experiments in which the beam energy and composition are very badly known.

In 1998 the experiment Super-Kamiokande (SK), located about 1000 m below the earth's surface in Japan, observed a difference between upward and downward going muon neutrinos. This difference could be attributed to neutrino oscillations which are only possible if neutrinos have mass. The incoming 'beam' of neutrinos was generated by interactions of primary cosmic rays within the earth's atmosphere. Its energy profile peaks at about 1 GeV but has long tails out to very high energies and also down to 0. This incoming energy uncertainty is obviously much larger than that in the ongoing LHC experiments. Still, this experiment gave the first evidence (together with the experiment at the Sudbury Neutrino Observatory (SNO), which concentrated on solar neutrinos) for BSM physics.

In the Super-Kamiokande (SK) experiment the difference between up going and down going muon neutrinos was attributed to a neutrino oscillation in which the muon neutrino changed its flavor into a tau neutrino over the distance $L$, i.e. essentially the earth's diameter. The oscillation probability from a muon (a) to a tau (b) neutrino is then given by (1)

$$P_{a \to b} = \sin^2 2\theta \, \sin^2 \left( \frac{1.27 \Delta m^2 (\text{eV}^2) L(\text{km})}{E_\nu (\text{GeV})} \right)$$

where $E_\nu$ is the energy of the incoming neutrino of flavor a and $\Delta m^2$ is the difference between the squared masses of the two neutrinos, $\theta$ is a mixing angle. One must note that this oscillation formula is not sensitive to the sign of $\Delta m^2$ nor to the absolute values of the neutrino masses.

Neutrinos come in three flavors. In that case the mixing matrix may not just contain additional mixing angles and mass differences, but also a CP violating phase $\delta_{CP}$ (2). For a detailed description of oscillations of three flavors one has to consider the effects of an interaction of the neutrinos with matter which affects the electron neutrinos, but not the other flavors simply because matter contains only electrons and no other leptons. The latter effect then provides a sensitivity not only to $\delta_{CP}$, but also to the ordering of neutrino masses. The so-called 'normal-hierarchy' (NH) is one in which the electron neutrino mass is lowest, the 'inverted-hierarchy' (IH) is the opposite. The most recent analyses of global neutrino oscillation data give a preliminary evidence for NH (2). No absolute scale for these masses can be obtained from oscillation experiments so that the lowest mass could still be zero.

While the SK experiment, together with the SNO experiment, proved the existence of neutrino masses and gave approximate values for the mixing angles and the mass differences later experiments tried to obtain more precise values for these quantities (2). Experiments have been set up at nuclear reactors with relatively short baselines to precisely determine some of the mixing parameters. In addition, so-called long baseline experiments are performed (3) in which a neutrino beam is produced at an accelerator lab where a high-intensity high-energy proton beam hits a nuclear target producing pions and kaons. These mesons then decay into leptons and neutrinos with broad energy distributions which propagate through the earth into a distant detector. Presently running are the experiments NOvA from Fermilab over a distance of about 810 km to a detector on the earth's surface and the T2K experiment in Japan that starts a beam at J-PARC and shoots it into the underground SK detector about 295 km away.

The 'flagship' experiment that should allow an ultimate answer both on the presence of CP violations in the weak interactions as well as on the mass hierarchy is the Deep Underground Neutrino Experiment (DUNE), a joint Fermilab-CERN experiment (4). Starting in about 2026 DUNE will shoot a beam from Fermilab through the earth into a deep-underground laboratory in the former Homestake Mine in South Dakota, about 1300 km away (see Fig. 1). This particle beam is very different from the one that nuclear physicists are used to. While beams in nuclear physics are of the order of micrometer wide at the ion source and at the target, here the neutrino beam is about 1 m wide at its origin at Fermilab and km wide at the detector in the gold mine in South Dakota. Its composition is not precisely known since it contains both muon- and electron-neutrinos and even antineutrinos. Furthermore, its energy distribution is wide. At DUNE it peaks at about 2 - 3 GeV and at T2K it peaks around 0.6 GeV.

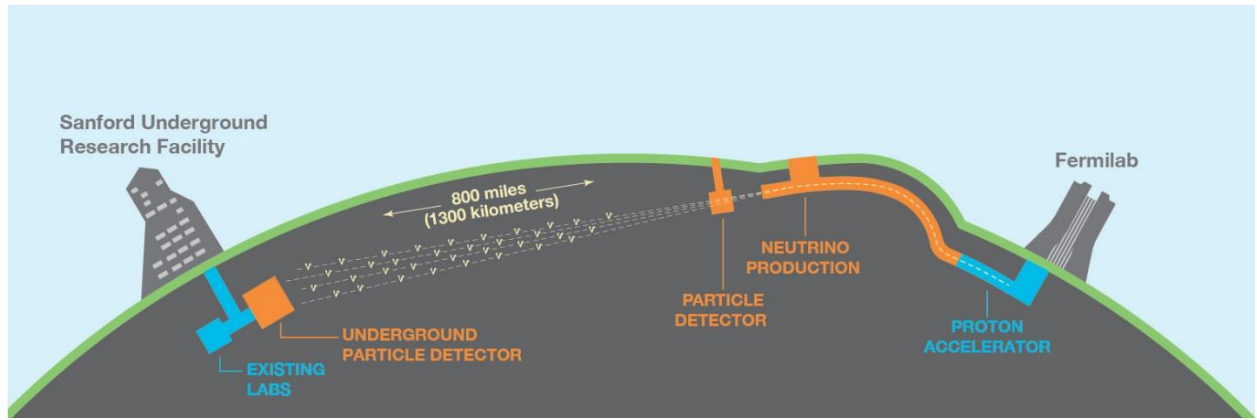

*Figure 1 Outline of the DUNE experiment with a beam starting at Fermilab, near Chicago, and going into a detector in the Sanford Underground Research Facility (SURF) in South Dakota*

This is a problem since the oscillation formula above (and its more refined version for 3 flavors) shows that the mass difference and the mixing parameter can be extracted only if the incoming neutrino beam energy is known. The energy distributions of two experiments are shown in Fig. 2 by the hatched areas. Fig. 2 also contains expected oscillation signals for the two mass hierarchies and various values of the CP-violating phase $\delta_{CP}$. These figures give a first impression on how well the neutrino energy has to be known in order to be able to distinguish between the different curves. For T2K this number is (very) roughly 50 MeV, for DUNE it is about 100 MeV.

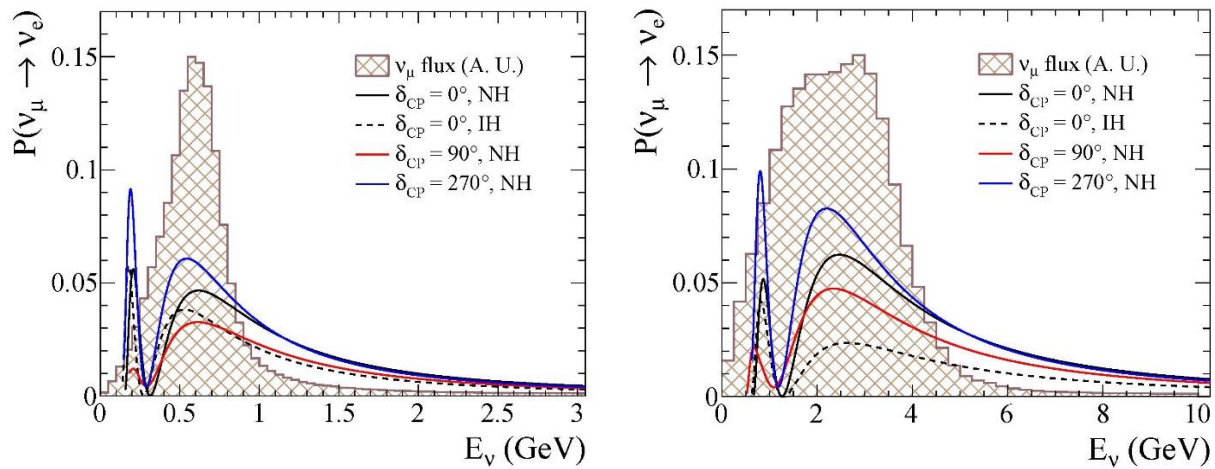

*Figure 2 Oscillation signals in T2K (left) and DUNE (right). The hatched areas give the neutrino energy-distributions for the two experiments. The various colored curves indicate how the oscillation signal depends on the mass hierarchies (NH and IH) as well as on a CP violating phase. Figure from M. Diwan, private communication.*

The neutrino energy must be reconstructed, event by event, from the observed final state. This is where Nuclear Physics comes in since all the detectors used nowadays are composed of nuclei. In the case of T2K this is water, for DUNE it will be liquid Argon. The final state of a neutrino-nucleus reaction is then determined not only by the initial, first neutrino-nucleon reaction, but also by the significant final state interactions in the nuclear target.

Two methods are commonly used for this energy reconstruction. The first one is the so-called 'kinematic method'. In it one uses the fact that in a Charged Current QuasiElastic (CCQE) scattering event on a neutron at rest the energy of the incoming neutrino is uniquely determined by the kinematics (energy and angle) of the outgoing lepton. In a nucleus the neutron is not at rest and not free, but instead moving with some momentum in a binding potential. This alone leads to an unavoidable smearing of the reconstructed energy around the true energy. More critical for a practical application of this method is that in a nuclear target the experimental identification of a true CCQE event as such is not possible. For example, an event in which first a pion was produced and then later on reabsorbed in the nuclear target cannot be distinguished from a true CCQE event. As a consequence, the energy thus determined cannot be the correct neutrino energy. Obviously, this complication is less important at the lower energies where the pion production probability is also low. T2K, with its beam energy peaked around 0.6 GeV, uses this method.

Fig. 3 illustrates the effects of the energy reconstruction for the T2K experiment and for the target nucleus $^{16}O$, the main ingredient of the water target. For this case a generator (GiBUU (5)) was used to first generate many millions of events with the T2K flux. For this event generation the 'true' neutrino energy is an input. These events were analyzed as if they were experimental data by using the kinematical method. The two upper curves show the event distributions measured in a near detector close to the neutrino source. The solid curve gives the distribution as a function of true energy, the dashed curve shows the distribution as a function of an energy that is reconstructed by the kinematical method. Obviously, the latter distribution is shifted downwards reflecting errors in the reconstruction.

These errors become essential at the far detector; the distributions there are shown by the lower 2 curves. Now the oscillatory pattern becomes washed out and the precise extraction of neutrino properties becomes very difficult.

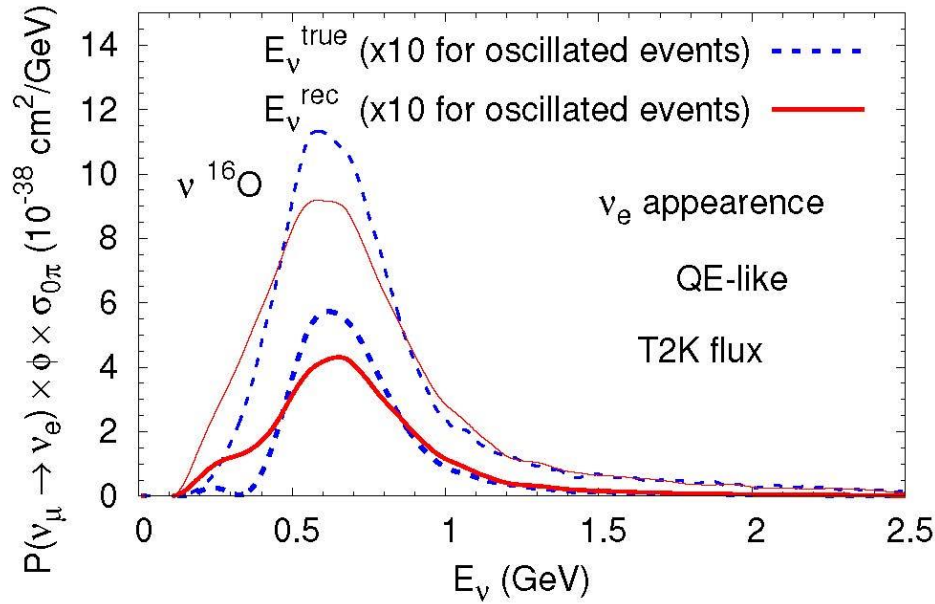

*Figure 3  Energy reconstruction of the oscillation signal in the T2K experiment. The two upper curves give the true (dashed) and reconstructed (solid) event distributions as a function of neutrino energy at a near detector. The two lower curves (magnified by a factor 10 for better. visibility) show the oscillation signal for electron neutrino appearance both for the true and the reconstructed distribution. The reconstruction was done  with the kinematical method. From Ref. (6)*

In higher-energy experiments such as DUNE about 2/3 of the total cross section goes into pion production, either through resonance excitations or through Deep Inelastic Scattering (DIS). For this reason there the so-called calorimetric method is preferred. In this method one tries to determine the beam energy by measuring the energies of all particles in the final state. This method sounds simple but is in reality also not straightforward because detectors have thresholds and efficiencies; often electrically neutral particles such as neutrons are not detected and the energy of the target remnant is not easy to determine.  The 'visible' energy is, therefore, always smaller than the true energy.

Thus both methods do not yield the correct neutrino energy right away. Both require an extrapolation from the determined energy to the true beam energy. This extrapolation is achieved by neutrino event generators. These generators describe the time-development of the whole reaction, from the very first, initial neutrino interaction with a moving and bound nucleon through all the final state interactions to the asymptotic state of free particles on their way to the detector. Imposing on the calculated final state the experimental restrictions, thresholds and efficiencies one can then calculate a so-called migration matrix that relates the actually measured final state energy to the true incoming energy (input into the generator calculation).

Creating such a generator is obviously a Nuclear Physics problem (7). In experiments at electron accelerator labs Nuclear Physics has dealt with the elementary processes and the final state interactions throughout the last few decades. Any theory of such processes has to be able to describe the first, initial interaction of the incoming electron with the nucleons in the nuclear target. Here all the reaction channels that also exist for a neutrino-nucleon reaction are present: quasielastic scattering, nucleon resonance excitation and – at the highest energy and four-momentum transfers – Deep Inelastic Scattering (DIS).

In addition, there are reaction processes that are typical for a nuclear target. Coherent pion production can take place; a process that has been observed with incoming photons at the accelerator MAMI. In addition to these coherent reaction channels there can be interactions in which the neutrino interacts with two nucleons simultaneously, so-called 2p-2h processes. In such reactions the two particles can be linked either by meson exchange currents or by short-range correlations.

The theoretical description of all these processes is complicated by the fact that the nucleons are bound inside the nucleus and that they are Fermi-moving. Thus, reaction amplitudes taken from collisions with individual free nucleons have to be properly extended into an off-shell regime. Electronuclear physics has dealt with such problems over the last 40 years and there is lots of theoretical knowledge about this available. In experimentalists' generators (8) the initial state is often simply described by a free Fermi gas, without any binding, and the outgoing state is again that of a free particle. However, there are more sophisticated methods available that start from a mean field potential and take RPA correlations in the excited state into account. Nuclear many-body theory can be used to calculate even more refined initial states by taking care of the short-range correlations between nucleons, leading to a nontrivial energy-momentum dispersion relation (spectral function) of the bound initial nucleon. A very good description of the inclusive QE process can also be achieved with the so-called scaling method in which the lepton-inclusive cross section is determined by a universal scaling function.

All these initial reaction processes have to be linked consistently to the final state interaction phase of the neutrino-nucleus collision. The particles that are present in the final state of any such first reaction traverse the nucleus before they can be detected. While doing so they experience final state interactions (fsi). They propagate in the same nuclear potential that was also active in the initial interaction; there is thus no exact factorization of initial and final state interactions. While they propagate they can experience all sorts of reactions: elastic and inelastic scattering, the latter often connected with pion production and also with charge transfer. The finally observed particles can, therefore, be quite different from the initially produced ones.

Essential for use in generators are, therefore, only those theories that can also provide the outgoing final states of the hadrons. Generators that are presently being used by experimentalists (8) do this by combining the first step with a simple Monte Carlo cascade in the target-density. In doing so they neglect any potentials and just correct the final energy by a binding energy parameter. This can, and must, obviously be improved by using state-of-the-art nuclear theory (7; 9).

Nuclear theory has gained a lot of experience in handling reactions of elementary particles on nuclei. For the present problem fully quantum mechanical methods cannot be used because they are not able to describe the full final state of the reaction. Instead, transport theory offers a state-of-the-art method for the description of such complex reactions. The foundations of it were laid about 35 years ago by G.F. Bertsch, P. Danielewicz and R. Malfliet, based on the even older Kadanoff-Baym equations for the non-equilibrium time development of a nuclear system. Transport theory allows to include a potential from the very first interaction through the final state interactions up to the very end of the reaction. It also provides the methods for off-shell transport; the latter becomes essential, for example, when a correlated nucleon with a collision-broadened spectral function leaves the nucleus and becomes 'sharp' again.

Transport codes were first used to describe heavy-ion collisions, where they still play an essential role. One of them, GiBUU (5), has also been extended to describe photo- and electro-nuclear processes. Because of this history the final state interactions in GiBUU have undergone a far-reaching test in many different reactions and kinematical regimes. It is then a small step to use the very same method also for the description of neutrino-nucleus reactions.

Presently running long-baseline experiments are located in Japan (T2K) and the USA (NOvA). The experiment DUNE will again be housed in the US. In addition, there are plans to build such an experiment attached to the European Spallation Source (ESS) in Sweden with a far detector at a distance of 300 – 500 km. The Neutrino Platform at CERN can act as a European focusing point for all these activities and plans.

While most of the experiments run overseas much of the underlying theory work for these neutrino-nucleus interactions is being done in Europe. The work started about 50 years ago; in 1972 a still relevant review article by Llewellyn-Smith appeared (10). While this early work relied on simplified models of nuclei about 30 years ago groups in Lyon and Torino and in the US extended it by using more sophisticated methods from nuclear structure theory. At present theory groups mainly in Belgium, Germany, Italy and Spain contribute significantly to the development of new methods to be used in generators. It is a lively scene where nuclear theory brings its own expertise into high energy physics labs and communities.

The theoretical developments by all these groups still have to make it into a new state-of-the-art neutrino event generator. Ultimately, the precision of extracted neutrino properties is determined by the quality of the generator. Generator development should, therefore, receive the same scrutiny (and funding support) as the experimental equipment.